\documentclass[iop]{emulateapj}
\usepackage{multirow}
\usepackage{bm}
\usepackage{amssymb}
\usepackage{amsmath}
\usepackage{latexsym}
\usepackage{graphicx}
\usepackage{ifsym}
\usepackage{bm}
\usepackage{tabularx}
\usepackage{hyperref}
\usepackage{enumitem}

\newcommand{\kms}{\mbox{km s$^{-1}$}}                                    

\slugcomment{}
\shorttitle{ALMA Dynamical Mass Estimate of FW Tau C}
\shortauthors{Wu \& Sheehan}

\begin{document}
\title{{\textbf {\large A\lowercase{n} ALMA D\lowercase{ynamical} M\lowercase{ass} E\lowercase{stimate of the} P\lowercase{roposed} P\lowercase{lanetary-mass} C\lowercase{ompanion} FW T\lowercase{au} C}}}

\author{Ya-Lin Wu \lowercase{and}
 Patrick D. Sheehan}
\affil{Steward Observatory, University of Arizona, Tucson, AZ 85721, USA\\{\it Accepted for Publication in ApJL}}

\begin{abstract}
Dynamical mass estimates down to the planet-mass regime can help to understand planet formation. We present Atacama Large Millimeter/submillimeter Array (ALMA) 1.3 mm observations of FW Tau C, a proposed $\sim$10 $M_{\rm Jup}$ planet-mass companion at $\sim$330 au from the host binary FW Tau AB. We spatially and spectrally resolve the accretion disk of FW Tau C in ${}^{12}$CO (2--1). By modeling the Keplerian rotation of gas, we derive a dynamical mass of $\sim$0.1 $M_\sun$. Therefore, FW Tau C is unlikely a planet, but rather a low-mass star with a highly inclined disk. This also suggests that FW Tau is a triple system consisting of three $\sim$0.1~$M_\sun$ stars. 
\end{abstract}

\keywords{accretion, accretion disks -- planets and satellites: individual (FW Tau C) -- stars: individual (FW Tau) -- techniques: interferometric}

\section*{\textbf {\normalsize1. I\lowercase{ntroduction}}}
In the past $\sim$10 years, many planet-mass companions at wide separations (few to tens of $M_{\rm Jup}$; tens to hundreds of astronomical units from host stars) have been discovered in direct imaging surveys. Their masses are usually determined by comparing observables like luminosity and effective temperature to predictions from evolutionary and atmospheric models (e.g., \citealt{C00,M07,SB12,Baraffe15}), which vary widely depending on the formation pathway (e.g., core accretion versus gravitational instability). As a result, it would be valuable if masses could be dynamically measured. As some of these substellar companions are young ($\lesssim$10 Myr) and have features associated with active accretion (e.g., \citealt{Z14}), their dynamical masses can be measured if their accretion disks can be spatially and spectrally resolved.

Among all of the known wide-separation companions, the proposed planet-mass object FW Tau C is the prime target to search for a Keplerian-rotating disk. FW Tau C is a tertiary companion located at $\sim$2\farcs3~projected separation ($\sim$330 au) to FW Tau AB, a close binary ($\sim$0\farcs08) of nearly equal-mass stars ($\sim$0.1~$M_\sun$) in the 2 Myr Taurus--Auriga star-forming region. FW Tau C was discovered by \cite{WG01} in their survey of binary stars, and its common proper motion was recently confirmed \citep{K14}. Studies have shown that FW Tau C has a rather flat near-infrared continuum owing to accretion-induced veiling, as well as many emission lines indicative of outflow and accretion activities \citep{WG01,B14}. The accretion disk was previously detected by the Atacama Large Millimeter/submillimeter Array (ALMA) in 1.3 mm dust continuum \citep{K15} and ${}^{12}$CO (2--1) \citep{C15}, and a 1--2 $M_\oplus$ of dust was inferred. Despite these observational efforts, FW Tau C's nature remains enigmatic because strong veiling inhibits accurate mass estimates. \cite{K14} derived a planetary mass of $10\pm4$~$M_{\rm Jup}$ from its dereddened $K'$ flux, while \cite{B14} suggested that FW Tau C could be a 0.03--0.15 $M_\sun$ brown dwarf or low-mass star embedded in an edge-on disk in order to explain its flat $K$-band spectrum and faint optical and near-infrared brightness.

Accurate mass measurement is therefore key to distinguishing the planet-mass scenario from the stellar-mass scenario. Here, we present the new ALMA 1.3 mm data from Cycle 3. With a $\sim$0\farcs2 beam and $\sim$0.3~\kms~velocity resolution, we spatially and spectrally resolve the gas disk and derive a dynamical mass of FW Tau C by modeling the Keplerian rotation. Parameters of the FW Tau system are summarized in Table \ref{tab:prop_fwtau}. 

\begin{deluxetable}{@{}lccl@{}}
\tablewidth{\linewidth}
\tablecaption{Properties of FW Tau \label{tab:prop_fwtau}}
\tablehead{
\colhead{\hspace{-20.5pt}Parameter} &
\colhead{\hspace{0pt}FW Tau AB} &
\colhead{\hspace{0pt}FW Tau C} &
\colhead{\hspace{0pt}References} 
}
\startdata \vspace{2pt}
Distance (pc)					& 	\multicolumn{2}{c}{$\sim$140}		& 	1		\\ \vspace{2pt}
Age (Myr)						& 	\multicolumn{2}{c}{$\sim$2}		& 	2		\\ \vspace{2pt}
Separation (\arcsec)				& 	\multicolumn{2}{c}{$\sim$2.3}		& 	3, 4		\\ \vspace{2pt}
PA (\degr)						& 	\multicolumn{2}{c}{$\sim$296}		&	3, 4		\\ \vspace{2pt}
SpT							&	M6~$\pm$~1					&  	$\cdots$ 	& 	3	\\ \vspace{2pt}
$A_V$ (mag)					&	$\sim$0.4						&	$\cdots$ 	&	3	\\ \vspace{2pt}
log($L/L_\sun$) 				&	$\sim$$-$1.1\tablenotemark{$a$}		& 	$\cdots$ 		&	5	\\ \vspace{2pt}
Mass ($M_\sun$)				&	$\sim$0.12\tablenotemark{$a$}		& 	$\sim$0.1		& 	5, 6
\enddata
\tablecomments{${}^a$ For each component.}
\tablerefs{(1) \cite{K94}, (2) \cite{K09}, (3) \cite{B14}, (4) \cite{K14}, (5) \cite{WG01}, (6) This work.}
\end{deluxetable}

\begin{figure*}
\centering
\includegraphics[angle=0,width=\linewidth]{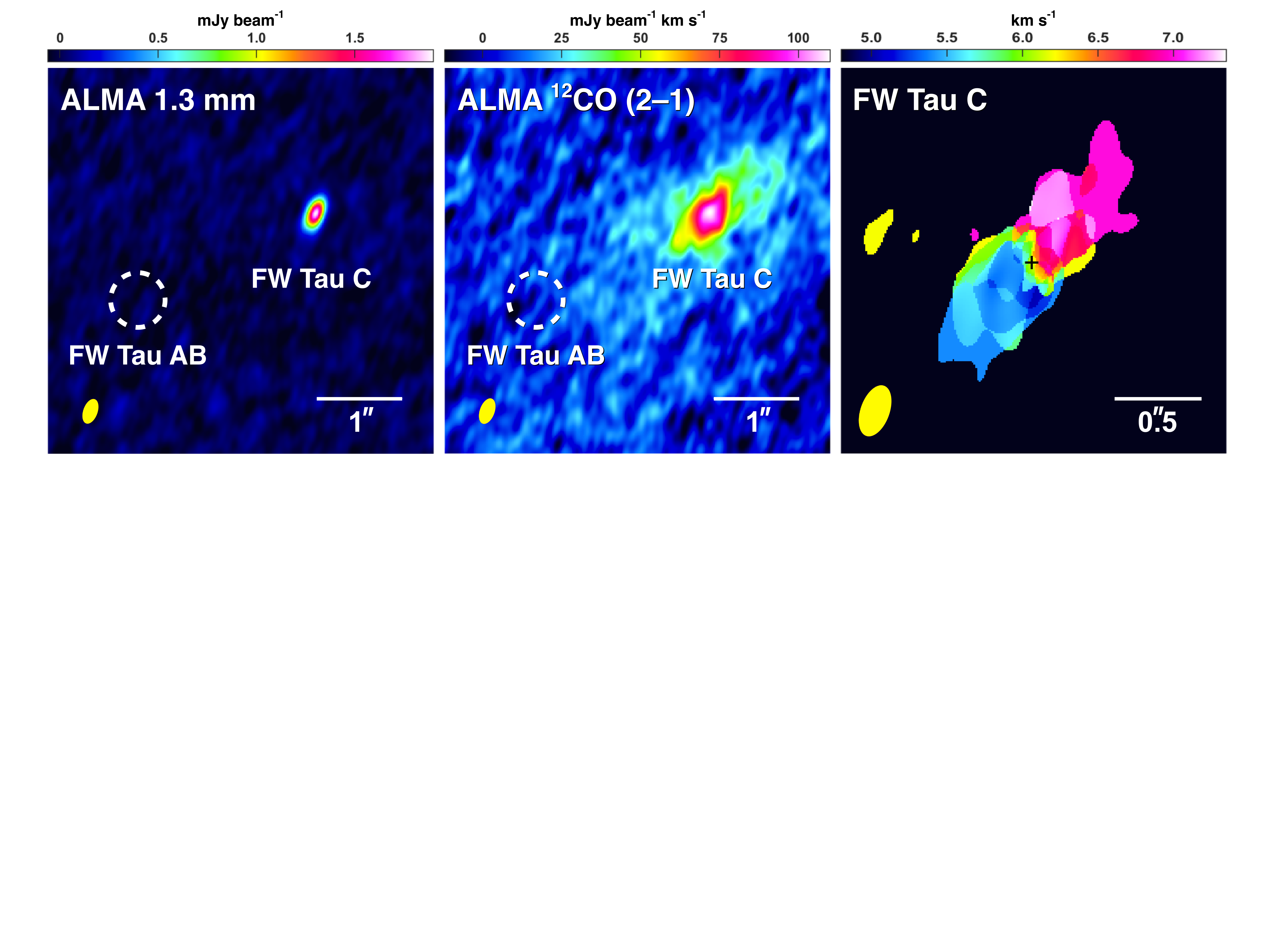}
\caption{Left: ALMA 1.3 mm dust continuum map of the FW Tau system. The primary stars FW Tau AB are not detected (dashed circle). Middle: ALMA ${}^{12}$CO (2--1) integrated intensity map (moment 0) shows a resolved accretion disk around FW Tau C. Right: the velocity field (moment 1) of FW Tau C's disk clearly shows a Keplerian rotation. The black cross marks the center of the disk in 1.3 mm continuum. Beam size $\sim$$0\farcs29\times0\farcs16$ and PA $\sim$160\degr. North is up and east is left.}
\label{fig:maps}
\end{figure*}

\begin{figure}
\centering
\includegraphics[angle=0,width=\linewidth]{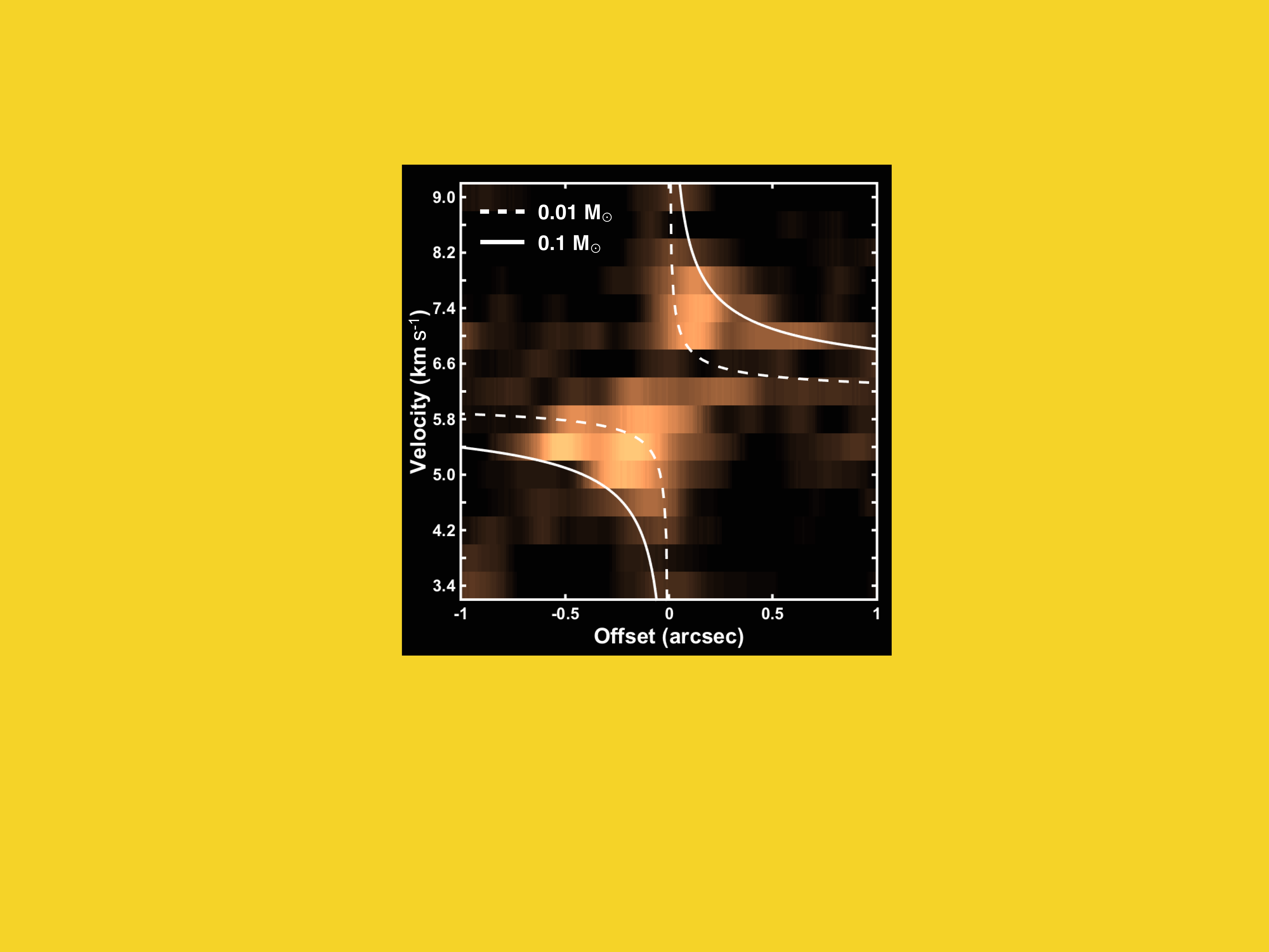}
\caption{Position--velocity diagram constructed along the major axis of the FW Tau C disk. Keplerian rotation curves for a 0.01~$M_\sun$ ($\sim$10 $M_{\rm Jup}$) planet and a 0.1~$M_\sun$ star are also plotted. FW Tau C is more consistent with a 0.1~$M_\sun$ star.}
\label{fig:pvdiagram}
\end{figure}

\section*{\textbf {\normalsize2. M\lowercase{ethodology}}}
\subsection*{2.1. O\lowercase{bservations and} D\lowercase{ata} R\lowercase{eduction}}
FW Tau was observed with ALMA Band 6 during Cycle 3, on 2016 September 14. During the observations thirty-six 12 m antennas were available, with baselines ranging from 15 to 3247 m. The Band 6 receiver was configured to have three basebands set up for dust continuum observations, centered at 233.0, 246.0, and 248.0 GHz and each with 2 GHz of bandwidth. The last baseband was configured with 3840 0.122 MHz channels centered at 230.538 GHz (0.32~\kms~velocity resolution; Hanning smoothed) in order to spatially and spectrally resolve ${}^{12}$CO (2--1) emission from the disk. J0510+1800 was used as the bandpass and flux calibrator, and J0433+2905 was used as the the gain calibrator. The on-source time was $\sim$13 minutes. 

The data were reduced using the ALMA pipeline in the \texttt{CASA} package. We employed one iteration of phase-only self-calibration to the 2 GHz continuum basebands. The baseband containing CO emission was much narrower, so it had much lower signal-to-noise ratio in continuum emission and we were unable to obtain good self-calibration solutions. We then CLEANed the calibrated data using the multi-frequency synthesis mode and natural weighting to enhance sensitivity in the images. The continuum map (left panel of Figure \ref{fig:maps}) has a beam size of $0\farcs29\times0\farcs15$ with a position angle (PA) of $160\fdg8$, and an rms of 35 $\mu$Jy beam$^{-1}$. The CO channel maps (top panel of Figure \ref{fig:channel_maps}) have a beam size of $0\farcs30\times0\farcs16$ with a PA of $160\fdg3$, and a mean rms of 4.5 mJy beam$^{-1}$ in signal-free channels. The integrated moment-zero and moment-one maps of the CO emission are shown in Figure \ref{fig:maps}.

\begin{deluxetable}{@{}lc@{}}
\tablewidth{\linewidth}
\tablecaption{FW Tau C Disk Properties\label{tab:disk_parameters}}
\tablehead{
\colhead{\hspace{-26pt}Parameter} &
\colhead{Value} 
}
\startdata \vspace{2pt}
$M_*$ ($M_\odot$) & $0.098 \pm 0.015$ \\ \vspace{2pt}
$M_{\text{disk,dust}}$ ($M_\earth$) & $1.15 \pm 0.01$  \\ \vspace{2pt}
$M_{\text{disk,gas}}$ ($M_\earth$) & $0.58\pm0.13$  \\ \vspace{2pt}
$R_{\text{disk}}$ (au)	&  \hspace{-8.5pt}$141.9 \pm 6.6$  \\ \vspace{2pt}
$\gamma$ &  $1.63 \pm 0.14$  \\ \vspace{2pt}
$T_0$ (K) &  $62.1 \pm 34.6$  \\ \vspace{2pt}
$q$ &  $0.14 \pm 0.10$  \\ \vspace{2pt}
$\xi$ (\kms) &  $0.65 \pm 0.08$  \\ \vspace{2pt}
$v_{\rm sys}$ (\kms)  & $6.10 \pm 0.02$ \\ \vspace{2pt}
$i$ (\degr) 	&   \hspace{-4.2pt}$62.8 \pm 2.4$  \\ \vspace{2pt}
PA (\degr)		&   \hspace{-4.2pt}$40.6 \pm 1.5$ 
\enddata
\label{tab:disk_parameters}
\end{deluxetable}

\subsection*{2.2. D\lowercase{isk} M\lowercase{odeling}: C\lowercase{ontinuum}}

The continuum emission from FW Tau C is, at best, only marginally resolved by our observations, so we use a simple geometrical model to fit the data. We assume that the continuum emission traces a uniform brightness disk, with a 1.3 mm brightness $F_{\nu}$, a radius $R_{\rm disk}$, and some inclination $i$ and position angle PA. We also allow the centroid of the emission to vary in our fit. We fit the model directly to the continuum visibilities using the Markov Chain Monte Carlo (MCMC) package \texttt{emcee} \citep{FM13}.

In order to measure the dust mass of the system, we assume that the continuum emission traces optically thin dust so that we can estimate the disk mass from
\begin{equation}
M_{\rm disk} = \frac{F_{\nu} \ D^2}{\kappa_{\nu} \, B_{\nu}(T)}
\end{equation}
\citep{B90}. We use standard assumptions of $T = 20$ K and $\kappa_{\nu} = 2.3$ cm$^2$ g$^{-1}$ and a distance to FW Tau C of 140 pc. 

\subsection*{2.3. D\lowercase{isk} M\lowercase{odeling}: K\lowercase{eplerian} R\lowercase{otation}}

Unlike the 1.3 mm continuum emission, FW Tau C's gas disk is well resolved by our ${}^{12}$CO (2--1) observations (see Figure \ref{fig:maps}), including a clear detection of spatially resolved Keplerian rotation. To model the data and determine disk and stellar parameters, we follow the modeling procedure described in \cite{Cz15} to fit our channel maps with synthetic channel maps produced from radiative transfer models. Such models can be used to measure disk parameters such as radius and inclination, as well as the stellar (or planetary) mass \citep{Cz15,Cz16}. Although we follow the procedure outlined by \citet{Cz15}, we have developed our own codes to run and fit these models to our data set.

\begin{figure*}
\centering
\includegraphics[angle=0,width=\linewidth]{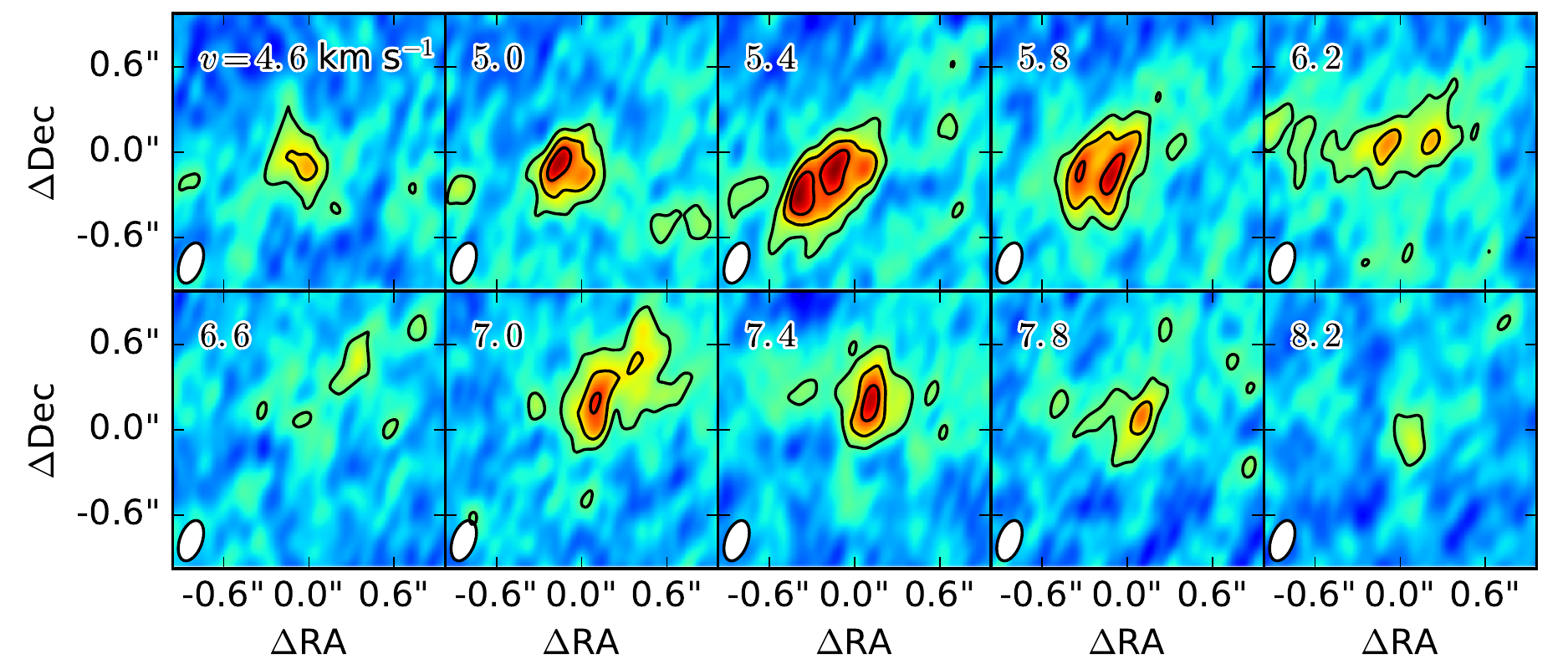}
\includegraphics[angle=0,width=\linewidth]{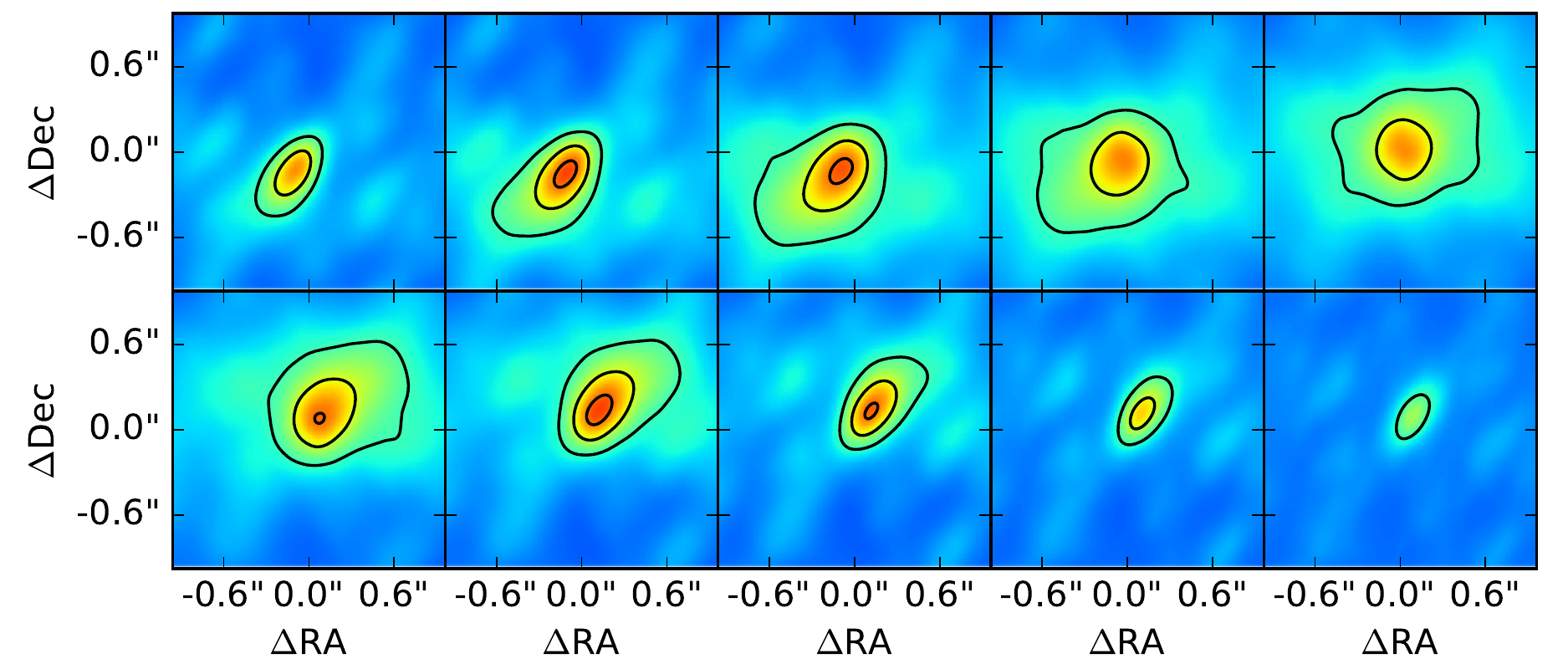}
\includegraphics[angle=0,width=\linewidth]{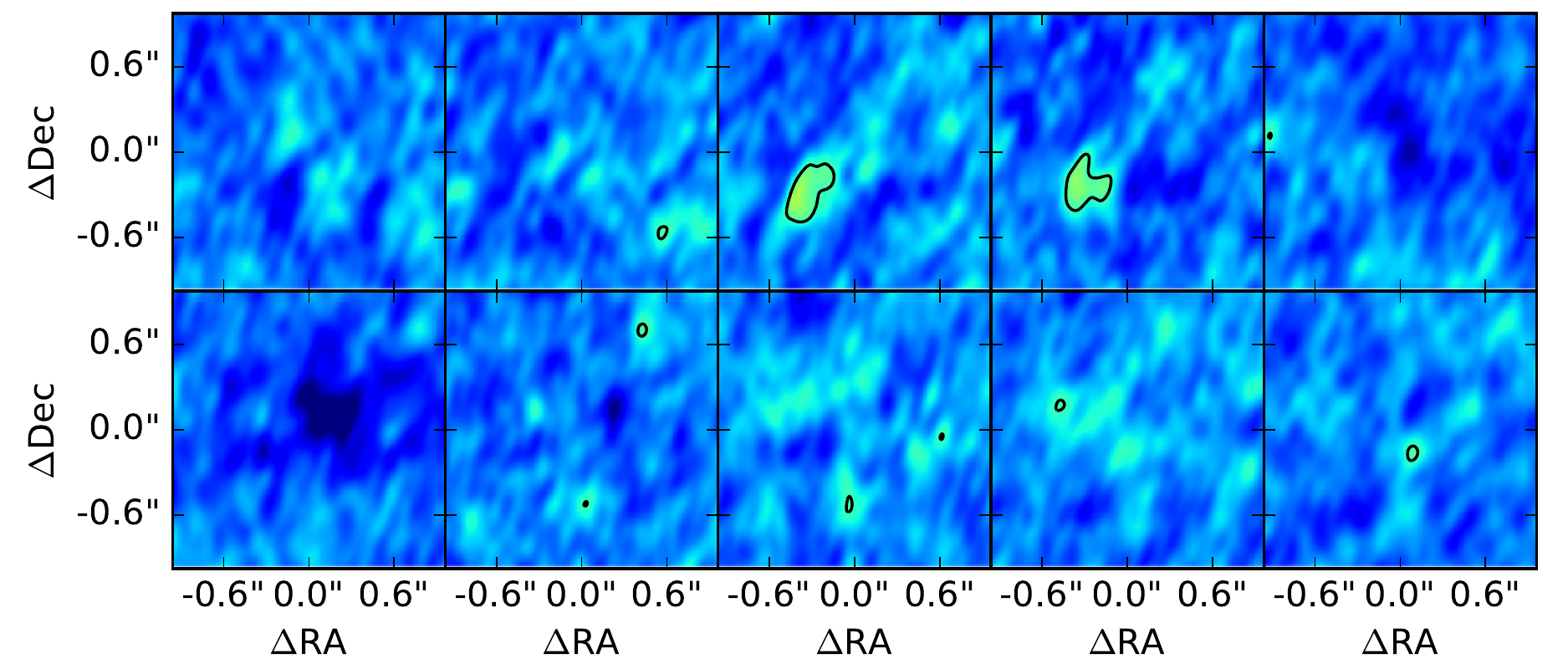}
\caption{We show our CO (2--1) channel maps (top) as well as the best-fit Keplerian disk model synthetic channel maps (center) and the residual channel maps (bottom). The model channel maps are made by sampling the model at the same baselines as the data and Fourier transforming to produce images. The residual channel maps are produced by subtracting the model visibilities from the visibility data and Fourier transforming to produce maps. The contours start at $3\sigma = 13.5$ mJy beam$^{-1}$ and are at intervals of $3\sigma$.}
\label{fig:channel_maps}
\end{figure*}

We assume that the ${}^{12}$CO (2--1) emission comes from a flared accretion disk, with a density profile described by
\begin{equation}
\rho(R,z) = \frac{\Sigma(R)}{\sqrt{2\pi} \, h(R)} \, \exp\left[-\frac{1}{2}\left(\frac{z}{h(R)}\right)^2\right],
\end{equation}
where $R$ and $z$ are defined in cylindrical coordinates, and $\Sigma(R)$ and $h(R)$ are the surface density and disk scale height, respectively. We assume that the disk has a power-law surface density profile: 
\begin{equation}
\Sigma(R) = \Sigma_0 \, \left(\frac{R}{R_0}\right)^{-\gamma},
\end{equation}
and we calculate the CO column density from this surface density profile as
\begin{equation}
N_{\rm CO}(R) = \frac{X_{\rm CO} \, \Sigma(R)}{\mu \, m_{\rm H}}.
\end{equation}
Here, $X_{\rm CO} = 1 \times 10^{-4}$ is the CO mass abundance fraction, and $\mu = 2.37$ is the mean molecular weight. The disk is truncated at an inner radius of 0.1 au and an outer radius of $R_{\rm disk}$, beyond which the density drops to zero.

We also assume that the disk is in local thermodynamic equilibrium and is vertically isothermal, with a radial temperature profile of 
\begin{equation}
T(R) = T_0 \, \left(\frac{R}{1~{\rm au}}\right)^{-q}.
\end{equation}
Under these assumptions, the scale height in the disk is set by the balance of thermal pressure and gravity such that
\begin{equation}
h(R) = \left( \frac{k_b \, R^3 \, T(R)}{G \, M_* \, \mu \, m_{\rm H}} \right)^{1/2},
\end{equation}
where $k_b$ is the Boltzmann constant, and $G$ is the gravitational constant.

Finally, rotation in the disk is assumed to be Keplerian, with an azimuthal velocity of
\begin{equation}
v_k = \sqrt{\frac{G \, M_*}{r}}.
\end{equation}
We assume that the velocities in the radial and vertical directions are zero. We also include microturbulent line broadening, which we assume is uniform throughout the disk, with a value of $\xi$ in units of \kms. Finally, we allow the star to have a systemic velocity, $v_{\rm sys}$, which Doppler shifts the velocity center away from zero.

In all, the density, temperature, and velocity structure of the system are described by the following parameters: $M_{\rm disk}$, $R_{\rm disk}$, $\gamma$, $T_0$, $q$, $\xi$, $M_*$, and $v_{\rm sys}$. We also allow the viewing geometry of the system, the inclination, and position angle to vary. In our model, the position angle of the disk is defined as the angle east of north of the projection of the disk angular momentum vector on the sky and ranges from $0^{\circ}$ to $360^{\circ}$. As our detection of CO (2--1) is not high sensitivity, we allow the inclination of the fit to range from $0^{\circ}$ to $90^{\circ}$. With higher sensitivity observations, however, it may be possible to distinguish between $i<90^{\circ}$ and $i>90^{\circ}$ \citep[e.g.,][]{R13,Cz15}. Initially, we center the CO data based on the results of our geometrical continuum modeling, but allow for a small deviation from that centering in our model. 

We fix the distance to the disk at 140 pc (see Table 1). The uncertainty in the distance, however, translates linearly into an uncertainty on the measured stellar mass \citep[e.g.,][]{Cz15,Cz16}, so we add the uncertainty on the distance estimate in quadrature with the uncertainty derived from our fit. Here, we assume a conservative distance uncertainty of 20 pc following \citet{Cz16}.

We use the 3D radiative transfer modeling package \texttt{RADMC-3D} \citep{Dullemond12} to calculate the level populations in each cell and produce synthetic ${}^{12}$CO (2--1) channel maps for a given set of model parameters. Those synthetic channel maps are Fourier transformed and fit directly to the visibilities using the MCMC fitting package \texttt{emcee} \citep{FM13}, with uniform priors for all parameters.

Although these models are computationally expensive, they can be run on powerful supercomputers so that the computations are spread out over a large number of central processing units (CPUs). For this particular instance, we run the fit over 128 CPUs on the University of Arizona El Gato supercomputer, and the modeling took a few days to converge. The models were determined to be converged when the \texttt{emcee} walkers had reached a steady state with measured best-fit values changing minimally over a large number of steps.

\section*{\textbf {\normalsize3. R\lowercase{esults}}}

Figure \ref{fig:maps} shows the 1.3 mm dust continuum, the ${}^{12}$CO (2--1) integrated intensity map (moment 0), and the intensity-weighted velocity map (moment 1) of the FW Tau C disk. Similar to \cite{K15} and \cite{C15}, no signal is found from the close binary FW Tau AB, and the $3\sigma$ upper limit for a unresolved source suggests a dust mass $\lesssim$0.07 $M_\oplus$. It is known that close binaries can shorten disk lifetimes (e.g., \citealt{Cieza09}), so the disk around FW Tau AB may be depleted already. For the companion FW Tau C, its dust disk is compact and likely unresolved; in contrast, the gas disk is more extended and clearly shows a Keplerian rotation.

Figure \ref{fig:pvdiagram} is the position--velocity diagram constructed along the major axis of the gas disk. On top of the PV diagram we also plot the Keplerian rotation curves for 10 $M_{\rm Jup}$ and 0.1 $M_\sun$ objects. It is clear that the velocity profile of the FW Tau C disk is incompatible with the 10 $M_{\rm Jup}$ rotation curve, but more consistent with that of a 0.1 $M_\sun$ star. Our disk modeling (see Figure \ref{fig:channel_maps} and Table \ref{tab:disk_parameters}) also indicates that FW Tau C's mass is $\sim$0.1~$M_\sun$, about 10 times higher than the 10 $M_{\rm Jup}$ suggested by \cite{K14}. Hence, FW Tau C is not a planetary-mass object, but is rather a low-mass star with an inclined disk, as suggested by \cite{B14}. The high mass of FW Tau C is also in agreement with some features in its spectrum that closely resemble T Tauri stars \citep{B14}. Our observations therefore suggest that FW Tau is a young triple system in which each of the stars has a mass of $\sim$0.1 $M_\sun$.

Searches for accretion disks in millimeter around directly imaged planet-mass companions have not yielded positive results (e.g., \citealt{I14,B15,M17,R17,Wolff17,W17}), although the disk around the free-floating planet OTS 44 has recently been imaged \citep{B17}. 
Current flux upper limits imply that these wide companion disks might have $<$0.1 $M_\oplus$ of dust, which in turn could imply a very short disk lifetime.
Future deep imaging down to a sublunar mass regime is perhaps needed to detect these disks, or place stringent constraints on their masses, and further elucidate the mass growth history of the planetary-mass companions at wide orbits. 

Finally, we summarize our disk modeling results as below and also in Table \ref{tab:disk_parameters}. Modeling the FW Tau C's dust disk as a circular Gaussian, we find that it has a 1.3 mm flux of $\sim$2.06 mJy, which equates to a dust mass of $\sim$1.15~$M_\oplus$ and is consistent with previous measurements in \cite{K15} and \cite{C15}. For the gas disk, our radiative transfer modeling, as demonstrated in Figure \ref{fig:channel_maps}, finds that the gas disk has a radius of $\sim$140 au, inclination of $\sim$63\degr, PA of $\sim$41\degr, systemic velocity of $\sim$6.1~\kms, and gas mass of $\sim$0.58~$M_\oplus$. Although recent studies have shown a low gas-to-dust ratio in protoplanetary disks (e.g., \citealt{WB14,A16}), we caution that with a single optically thick CO line, we are very likely to underestimate the true gas mass by a factor of few or even orders of magnitude (e.g., \citealt{Y17}). As \cite{Y17} pointed out, to accurately derive the gas mass, one should observe low-transition lines to ameliorate the optical-depth effects and should also use multiple CO isotopologues to characterize the chemical depletion and abundance variation across the disk.

\section*{\textbf {\normalsize4. S\lowercase{ummary}}}
FW Tau C is a wide-orbit companion at $\sim$330 au from the close binary FW Tau AB that has been suggested to have a planet-like mass of 10 $M_{\rm Jup}$. Here, we have used ALMA to detect FW Tau C's accretion disk in a 1.3 mm dust continuum and ${}^{12}$CO (2--1) emission. We find that the gas motion is both spatially and spectrally resolved and clearly follows Keplerian rotation, enabling a dynamical mass estimate of the central object. We show that FW Tau C's mass is in fact $\sim$0.1 $M_\sun$, so it is more likely a low-mass star embedded in an inclined accretion disk.

\acknowledgements
We thank the referee for very helpful comments. We thank Laird Close, Johanna Teske, Yu-Cian Hong, and Jing-Hua Lin for discussions. Y.-L.W. is supported by the NASA Origins of Solar Systems award and the TRIF fellowship. This Letter makes use of the following ALMA data: ADS/JAO.ALMA\#2015.1.00773.S. ALMA is a partnership of ESO (representing its member states), NSF (USA) and NINS (Japan), together with NRC (Canada), NSC and ASIAA (Taiwan), and KASI (Republic of Korea), in cooperation with the Republic of Chile. The Joint ALMA Observatory is operated by ESO, AUI/NRAO, and NAOJ. The National Radio Astronomy Observatory is a facility of the National Science Foundation operated under cooperative agreement by Associated Universities, Inc.


\begin{thebibliography}{}
\bibitem[Andrews et al.(2013)]{A13}Andrews, S. M., Rosenfeld, K. A., Kraus, A. L., \& Wilner, D. J. 2013, \apj, 771, 129
\bibitem[Ansdell et al.(2016)]{A16}Ansdell, M., Williams, J. P., van der Marel, N., et al. 2016, \apj, 828, 46
\bibitem[Baraffe et al.(2015)]{Baraffe15}Baraffe, I., Homeier, D., Allard, F., \& Chabrier, G. 2015, \aap, 577, A42
\bibitem[Bayo et al.(2017)]{B17}Bayo, A., Joergens, V., Liu, Y., et al. 2017, \apjl, 841, L11
\bibitem[Beckwith et al.(1990)]{B90}Beckwith, S. V. W., Sargent, A. I., Chini, R. S., \& Guesten, R. 1990, \aj, 99, 924
\bibitem[Bowler et al.(2015)]{B15}Bowler, B. P., Andrews, S. M., Kraus, A. L., et al. 2015, \apjl, 805, L17
\bibitem[Bowler et al.(2014)]{B14}Bowler, B. P., Liu, M. C., Kraus, A. L., \& Mann, A. W. 2014, \apj, 784, 65
\bibitem[Bowler et al.(2011)]{Bowler11}Bowler, B. P., Liu, M. C., Kraus, A. L., Mann, A. W., \& Ireland, M. J. 2011, \apj, 743, 148
\bibitem[Caceres et al.(2015)]{C15}Caceres, C., Hardy, A., Schreiber, M. R., et al. 2015, \apjl, 806, L22
\bibitem[Chabrier et al.(2000)]{C00}Chabrier, G., Baraffe, I., Allard, F., \& Hauschildt, P. 2000, \apj, 542, 464
\bibitem[Cieza et al.(2009)]{Cieza09}Cieza, L. A., Padgett, D. L., Allen, L. E., et al. 2009, \apjl, 696, L84
\bibitem[Czekala et al.(2015)]{Cz15}Czekala, I., Andrews, S. M., Jensen, E. L. N., et al. 2015, \apj, 806, 154
\bibitem[Czekala et al.(2016)]{Cz16}Czekala, I., Andrews, S. M., Torres, G., et al. 2016, \apj, 818, 156
\bibitem[Dullemond(2012)]{Dullemond12}Dullemond, C. P. 2012, RADMC-3D: A Multi-purpose Radiative Transfer Tool, Astrophysics Source Code Library, ascl:1202.015
\bibitem[Foreman-Mackey et al.(2013)]{FM13}Foreman-Mackey, D., Hogg, D. W., Lang, D., \& Goodman, J. 2013, \pasp, 125, 306
\bibitem[Isella et al.(2014)]{I14}Isella, A., Chandler, C. J., Carpenter, J. M., P\'{e}rez, L. M., \& Ricci, L. 2014, \apj, 788, 129
\bibitem[Kenyon et al.(1994)]{K94}Kenyon, S. J., Dobrzycka, D., \& Hartmann, L. W. 1994, \aj, 108, 1872
\bibitem[Kraus et al.(2015)]{K15}Kraus, A. L., Andrews, S. M., Bowler, B. P., et al. 2015, \apjl, 798, L23
\bibitem[Kraus \& Hillenbrand(2009)]{K09}Kraus, A. L., \& Hillenbrand, L. A. 2009, \apj, 704, 531
\bibitem[Kraus et al.(2014)]{K14}Kraus, A. L., Ireland, M. J., Cieza, L. A., et al. 2014, \apj, 781, 20
\bibitem[Lynden-Bell \& Pringle(1974)]{LBP74}Lynden-Bell, D., \& Pringle, J. E. 1974, \mnras, 168, 603
\bibitem[MacGregor et al.(2017)]{M17}MacGregor, M. A., Wilner, D. J., Czekala, I., et al. 2017, \apj, 835, 17
\bibitem[Marley et al.(2007)]{M07}Marley, M. S., Fortney, J. J., Hubickyj, O., Bodenheimer, P., \& Lissauer, J. J. 2007, \apj, 655, 541
\bibitem[Ricci et al.(2017)]{R17}Ricci, L., Cazzoletti, P., Czekala, I., et al. 2017, \aj, 154, 24
\bibitem[Rosenfeld et al. (2013)]{R13}Rosenfeld, K. A., Andrews, S. M., Hughes, A. M., et al. 2013, \apj, 774, 16
\bibitem[Spiegel \& Burrows(2012)]{SB12}Spiegel, D. S., \& Burrows, A. 2012, \apj, 745, 174
\bibitem[White \& Ghez(2001)]{WG01}White, R. J., \& Ghez, A. M. 2001, \apj, 556, 265
\bibitem[Williams \& Best(2014)]{WB14}Williams, J. P., \& Best, W. M. J. 2014, \apj, 788, 59
\bibitem[Wolff et al.(2017)]{Wolff17}Wolff, S. G., M\'{e}nard, F., Caceres, C., et al. 2017, \aj, 154, 26
\bibitem[Wu et al.(2017)]{W17}Wu, Y.-L., Sheehan, P. D., Males, J. R., et al. 2017, \apj, 836, 223
\bibitem[Yu et al.(2017)]{Y17}Yu, M., Evans, N. J., Dodson-Robinson, S. E., Willacy, K., \& Turner, N. J. 2017, \apj, 841, 39
\bibitem[Zhou et al.(2014)]{Z14}Zhou, Y., Herczeg, G. J., Kraus, A. L., Metchev, S., \& Cruz, K. L. 2014, \apjl, 783, L17
\end{thebibliography}
\end{document}